\documentclass{vgtc}                          

\ifpdf%
  \pdfoutput=1\relax                   
  \usepackage{graphicx}                
    
  \DeclareGraphicsExtensions{.pdf,.png,.jpg,.jpeg} 
\else%
  \usepackage{graphicx}                
  \DeclareGraphicsExtensions{.eps}    
\fi%

\graphicspath{{figures/}{pictures/}{images/}{./}} 

\usepackage{microtype}                 
\PassOptionsToPackage{warn}{textcomp}  
\usepackage{textcomp}                  
\usepackage{mathptmx}                  
\usepackage{times}                     
         
\usepackage{cite}                      
\usepackage{tabu}                      
\usepackage{booktabs}                  

\usepackage{xspace}
\usepackage{float}
\usepackage{subfig}
\usepackage{todonotes}

\onlineid{0}

\vgtccategory{Research}

\vgtcinsertpkg

\preprinttext{}

\title{Design, Assembly, Calibration, and Measurement of an Augmented Reality Haploscope}

\author{Nate Phillips\thanks{e-mail: Nathaniel.C.Phillips@ieee.org}\\ %
        \scriptsize Mississippi State University %
\and Kristen Massey\thanks{e-mail: kristenmassey@ieee.org}\\ %
     \scriptsize Mississippi State University %
\and Mohammed Safayet Arefin\thanks{e-mail:arefin@acm.org }\\ %
     \scriptsize Mississippi State University
\and J. Edward Swan II\thanks{e-mail: swan@acm.org}\\ %
     \scriptsize Mississippi State University %
     }

\teaser{
  \includegraphics[width=6.8in]{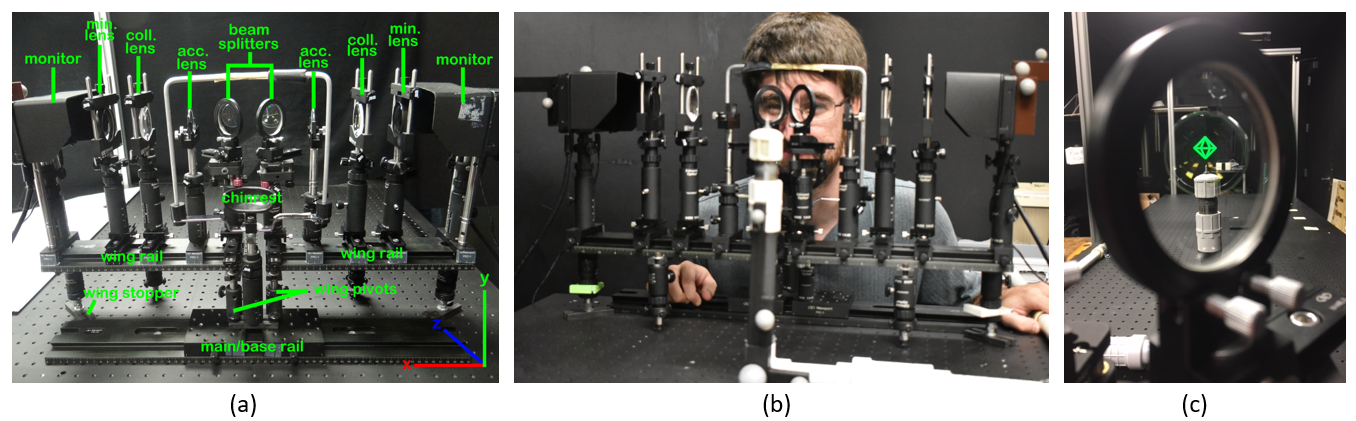}
  \caption{The Augmented Reality (AR) haploscope. (a) A front view of the AR haploscope with labeled components \cite{Massey:2018:PCOH}. (b) A user participating in perceptual experiments. (c) The user's view through the haploscope.  With finely adjusted parameters such as brightness, binocular parallax, and focal demand, this object appears to have a definite spatial location.}
  \label{f:haploscope}
}

\abstract{%
A haploscope is an optical system which produces a carefully controlled virtual image.  Since the development of Wheatstone's original stereoscope in 1838, haploscopes have been used to measure perceptual properties of human stereoscopic vision. 
This paper presents an augmented reality (AR) haploscope, which allows the viewing of virtual objects superimposed against the real world.  
Our lab has used generations of this device to make a careful series of perceptual measurements of AR phenomena, which have been described in publications over the previous 8 years.  
This paper systematically describes the design, assembly, calibration, and measurement of our AR haploscope.  
These methods have been developed and improved in our lab over the past 10 years.  
Despite the fact that 180 years have elapsed since the original report of Wheatstone's stereoscope, we have not previously found a paper that describes these kinds of details.
}

\begin{document}
\maketitle
\section{Introduction}
Mixed reality has been an active field of research for the past 50 years~\cite{Sutherland:1968}, but recent advances and interest have dramatically accelerated developments in the field.  This has resulted, lately, in an explosive increase in the development of virtual and augmented reality (AR) display devices, such as the Vuzix STAR, Oculus Rift, Google Glass, Microsoft HoloLens, HTC Vive, Meta 2, and Magic Leap One.  These displays have inspired consumer and business interest, increased demand for VR and AR applications, and motivated increased investment in VR and AR development and research~\cite{Shirer:2018}.  

This investment and development is stymied by an incomplete knowledge of key underlying research.  Among this unfinished research, our lab has focused on addressing questions of AR perception; in order to accomplish the ambitious goals of business and industry, a deeper understanding of the perceptual phenomena underlying AR is needed \cite{Kruijff:2010}.  All current commercial AR displays have certain limitations, including fixed focal distances, limited fields of view, non-adjustable optical designs, and limited luminance ranges, among others.  These limitations hinder the ability of our field to ask certain research questions, especially in the area of AR perception. 

Therefore, our lab has developed a custom AR display (Figure~\ref{f:haploscope}), which we call the \emph{AR Haploscope}, assembled from off-the-shelf optical components.  Our design, iterated through several generations and research projects\cite{Cook:2018, Hua:2014, singh2010depth, Singh:2013, Singh:2017, singh2018effect}, is based on previous haploscope research (e.g., \cite{Williams:1959,NASA:1975}), which has been widely used in the field of visual perception~\cite{Westheimer:2006}.  A \emph{haploscope} is an optical system that produces tightly-controlled virtual images, typically with controlled accommodative demand, presented angle, brightness, divergence, and image choice~\cite{Hua:2014, Cook:2018, Singh:2017}.  Such a system is highly controllable, can be adjusted for different inter-pupillary distances, can be used in a wide range of experiments, and can be reliably re-used.  

The advantages of using a haploscope come with the additional burden of calibration.  Through our own experience, we have discovered that the calibration of an AR haploscope is a non-trivial task.  There are several important factors to consider, as well as many potential pitfalls \cite{Rolland:2005}. The difficulty is compounded by a general dearth of published research on the topic of haploscope calibration; the authors have looked for, but not found, such a publication.  

Therefore, this paper seeks to contribute a systematic description and evaluation of the design, assembly, and calibration of an AR haploscope system.

\section{Background}

Since the discovery of perspective in the 14th century by Italian Renaissance painters and architects, scholars and scientists have been studying human stereo vision.  In 1838, Wheatstone~\cite{Wheatstone:1838} gave the first scientific description of the phenomena of stereopsis.  He also described the \emph{stereoscope}---the seminal instrument for precisely displaying a stereo pair of images.  A haploscope is fundamentally a stereoscope that has been  adapted for laboratory use.

Historically, haploscopes are instruments that are used to observe stereo vision and measure accommodation, vergence, and other properties of human eyes, across many disparate research fields. In 1959, Williams~\cite{Williams:1959} designed and developed a haploscope to study accommodation and convergence. NASA~\cite{NASA:1975} developed the Baylor Mark III Haploscope to measure, record and analyze the binocular vision of astronauts during spaceflight.  Ellis et al.~\cite{ellis1994distance}, in 1994, produced a novel head-mounted haploscope system to study the effects of interposition and occlusion on virtual images.  Finally, Rolland et al.~\cite{Rolland:2005} developed and used their own haploscope in order to study the accuracy and precision of depth judgments. These examples represent only a small sampling of historical haploscope usage, but are largely representative of haploscopes in research. 

In augmented reality in particular, haploscopes are useful for investigating topics like IPD mismatch, accommodation-vergence mismatch, depth perception, and various other important research areas. In order to analyze near field depth perception in AR, Singh designed and built a haploscope with dynamic focus adjustment and rotatable optics~\cite{Singh:2013, Swan:2015, Singh:2017}. Another haploscope system was created by Banks et al.~\cite{banks} to examine binocular disparity and eye-position. In addition, Domini~\cite{brown, Brown:2008} created a haploscope setup in order to analyze vergence angle effects.

\section{Haploscope Design}

\begin{figure}[tb]
 \centering
 \includegraphics[width=\columnwidth]{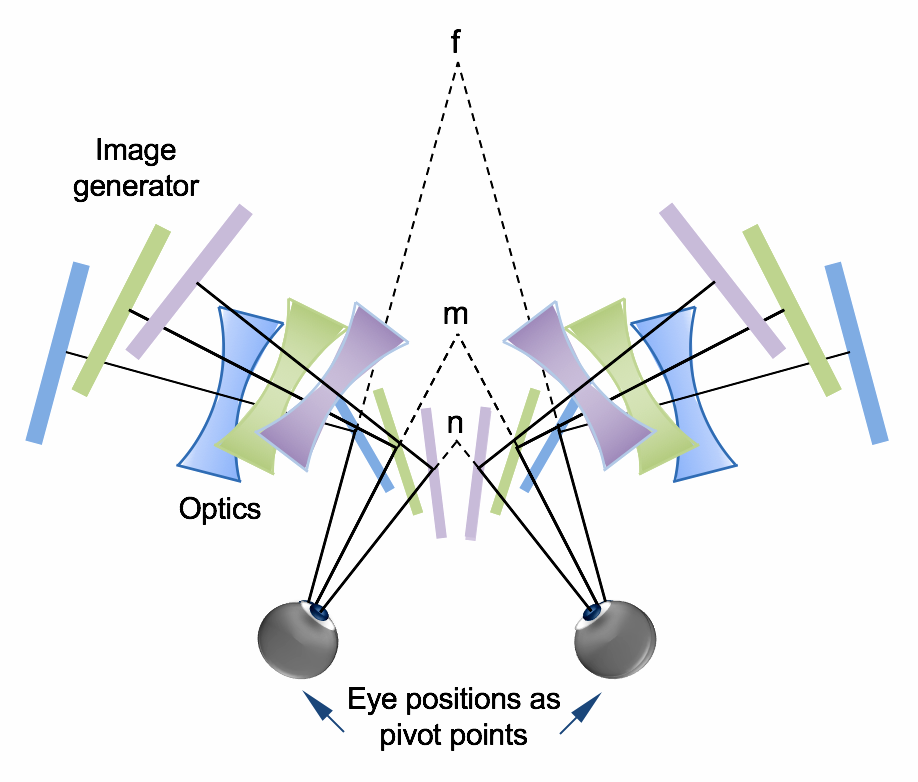}
 \caption{An example of how the haploscope wings rotate to match different focal distances, from Singh \cite[Figure 4.7]{Singh:2013}.  The displays rotate inward as a user's eyes rotate, ensuring that the user maintains a view fixed at the optical center where distortion is minimized.}
 \label{fig:gurjot-pivots}
\end{figure}

\begin{figure*}
\includegraphics[width=\linewidth]{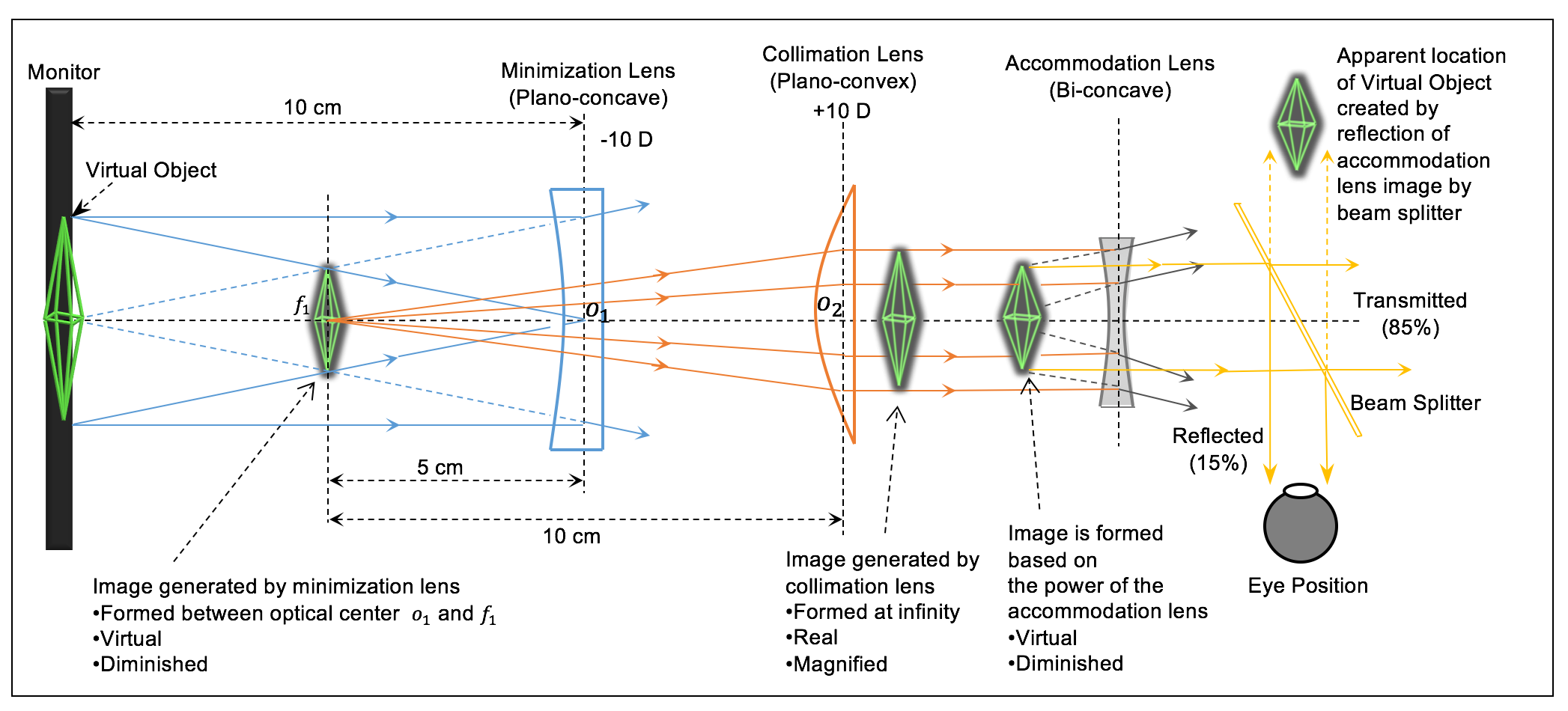}
\caption{Ray diagram of the side of the haploscope, based on Singh~\cite[Figure 4.4]{Singh:2013}.  This diagram showcases the path of the virtual object as it is generated by the monitor, shrunk by the minimization lens, collimated by the collimating lens, set to a specific focal distance by the accommodation lens, and finally reflected directly into a user's eye by the beamsplitter.}
\label{fig:ray-diagram}
\end{figure*}

A haploscope is an augmented reality tabletop apparatus that presents images to users through a lens system (Figure~\ref{f:haploscope}). Our particular haploscope, which is based on Singh's design~\cite{singh2018effect, Singh:2013}, presents an image and then collimates it.  After collimation, the \textit{accommodation lens} forms an image at any arbitrary distance.  This apparatus also allows free rotation about the modeled eye position of the user, allowing $\alpha$, sometimes also called the vergence angle or angle of binocular parallax, to be adjusted without adding optical distortion~\cite{singh2018effect}.

It may be instructive to examine the workings of the haploscope, outlined in brief above, in slightly more detail (Figure~\ref{fig:ray-diagram}).  At the start, the haploscope monitor generates an image source.  Then, that image is shrunk by the \textit{minimization lens}, a -10 diopter concave lens, producing a minified virtual image 5 cm in front of the lens.  This minified image, in turn, is collimated by the \textit{collimating lens}, a 10 diopter convex lens, positioned 10 cm in front of the minified image.  Now, collimated light, definitionally, is broadcast out in purely parallel rays of light, formed at optical infinity.  Once the collimated image source nears the user's eye, it is focused, by application of a particular negative power concave lens, to the desired focal distance.  Finally, the output of this lens system is reflected into a user's eye by a \textit{beamsplitter}, a partially reflective piece of glass.  This beamsplitter allows the user to see both the reflected virtual object and the real world displayed beyond the beamsplitter~\cite{singh2018effect}.

However, even this is not enough to fully control all the cues that a user needs to perceive an object.  Most notably, users also need to be able to converge appropriately to the visual target.  During vergence, a user's eyes rotate inward or outward so that they can center the viewed object in their vision.  For objects at a specific distance, this effect causes the eyes to rotate to a particular angle, $\alpha$, which can be readily calculated by the formula:
\begin{equation}
\alpha  = \arctan{\frac{\textrm{object distance}}{(\textrm{IPD/2})}}\label{eq:convergence},
\end{equation}
where IPD is the user's interpupillary distance.  

As such, this haploscope design needs to be able to accurately and precisely rotate to a variety of vergence angles.  With a haploscope, like ours, that rotates about the modeled eye position, the system's optical axes can be rotated to be collinear with the user's optical axes, eliminating this additional distortion (Figure~\ref{fig:gurjot-pivots}). 

At this point, it may be instructive to briefly discuss our assumptions about human vision.  We assume that a human eye can be accurately modeled as a simple schematic eye with a center of rotation, a single nodal point, and a pupil, all of which are co-linear with each other~\cite{Westheimer:2006, jones2016schematic}.  This means, in practice, that human vision takes place on the axis that includes the pupil, the nodal point (where human visual perception originates), and the center of eye rotation.  This model turns out to be rather important, as it has particularly key implications for how we set up the haploscope for a given user.

First, we measure user IPD at optical infinity, when his or her gaze vectors are essentially parallel.  This distance is important primarily because it is how far the center of a user's eyes are from each other; thus, adjusting the haploscope to this position allows us to align the centers of rotation of a user's eyes with the optical system.

Since the centers of the user's eyes and the haploscope system's centers of rotation are coincident, the haploscope optical axes remain aligned with the user's optical axes, even as the eyes/haploscope rails rotate.  This means that each point on the user's optical axes, including the nodal point, is presented with the same optical stimuli, regardless of rotation.  

\section{Assembly and Calibration}

\begin{figure}[t]%
 \centering
 \includegraphics[width=\columnwidth]{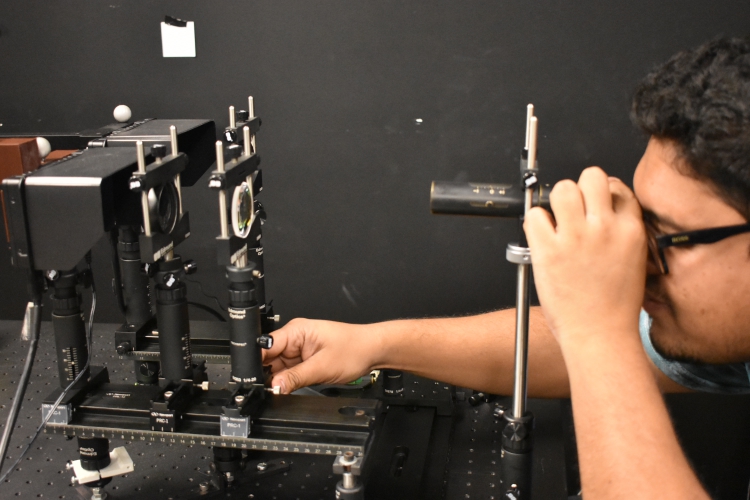}
 \caption{This figure shows an experimenter adjusting the minimization and collimation lenses during dioptometer calibration.  This calibration step is important to eliminate focal blur and to verify that the monitor image is appropriately collimated before entering the rest of the lens system.}%
 \label{f:Diopter}%
\end{figure}

\captionsetup[subfigure]{subrefformat=simple,labelformat=simple,listofformat=subsimple}
\renewcommand\thesubfigure{(\alph{subfigure})}

\begin{figure}[t]%
 \centering 
 \subfloat[Monitor adjustment]{
    \includegraphics[width=.44\columnwidth]{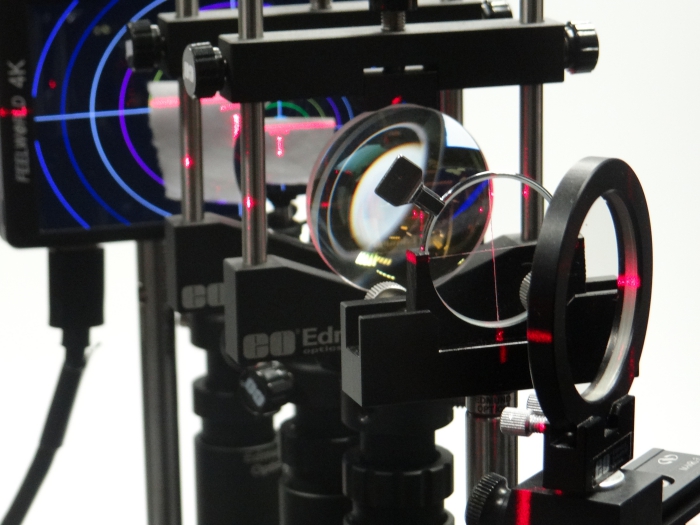}
    \label{f:IPDMonitor}%
  }
 \subfloat[Straight view]{
   \includegraphics[width=.44\columnwidth]{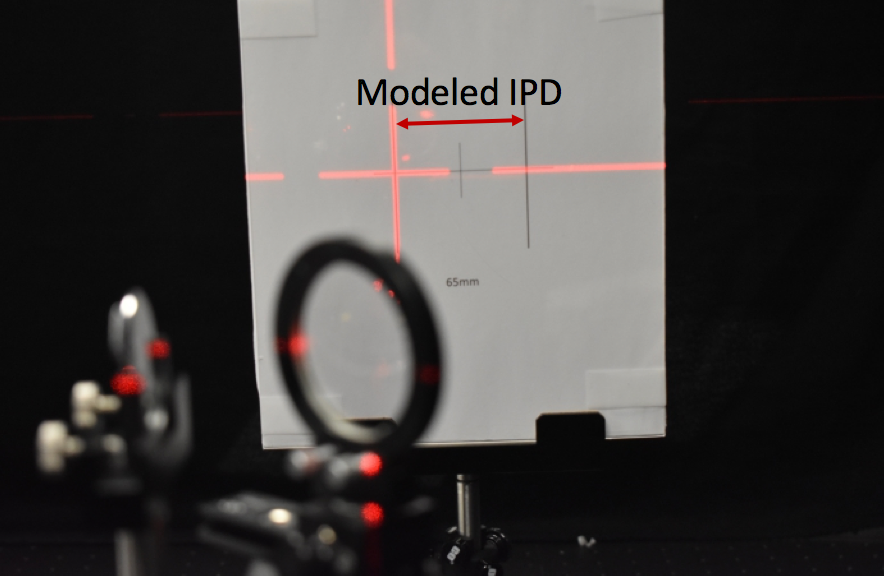}
    \label{f:IPDCalibrationTarget}%
  }
  \qquad
 \subfloat[Calibration target adjustment]{
    \includegraphics[width=.44\columnwidth]{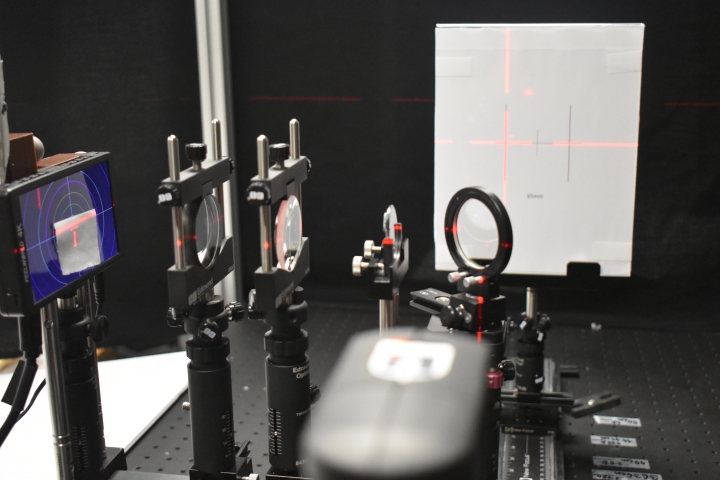}
    \label{f:IPDTotal}%
  }
 \subfloat[Overall]{
   \includegraphics[width=.44\columnwidth]{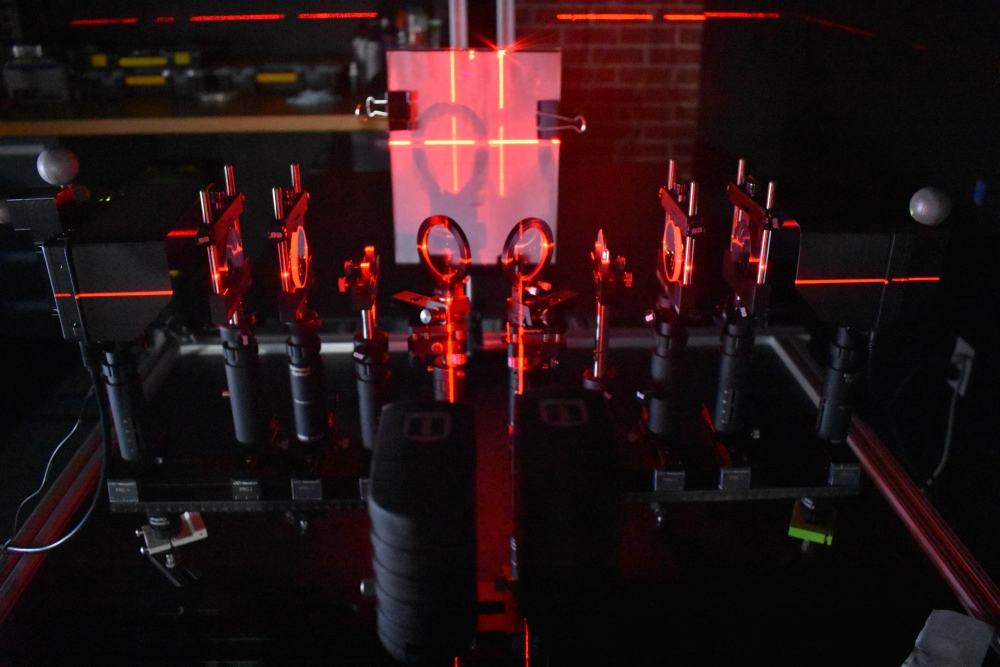}
    \label{f:IPDBigPicture}%
  }
 \caption{IPD Calibration for the left wing, with varying areas in focus: (a) monitor/laser level centering; (b) calibration target centering; (c) an overview of the calibration system; and (d) fully aligned optical axes centered with respect to the calibration target.  This alignment verifies that both rails' optical axes are coincident with the optical axes of a user with a given IPD.}%
 \label{f:GazeCond}%
\end{figure}

\begin{figure}[t]%
 \centering
 \includegraphics[width=\columnwidth]{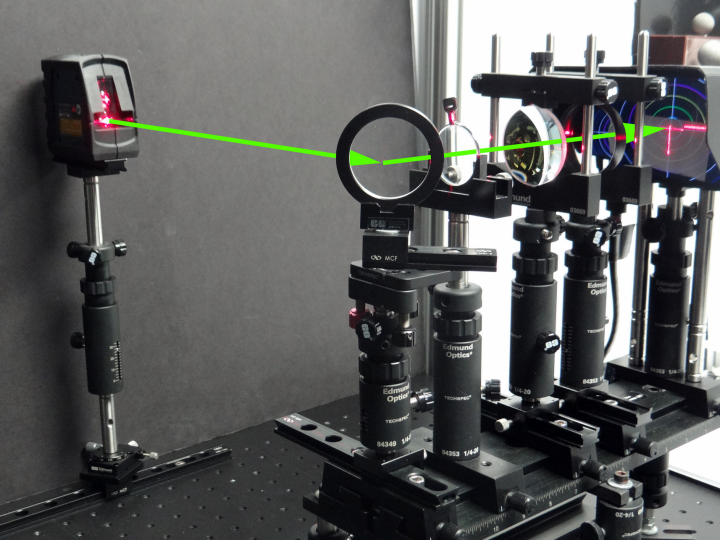}
 \caption{In beamsplitter calibration, the laser level passes through all optical components and rests centered on the monitor crosshairs, reducing optical distortion\cite{Massey:2018:PCOH}.}%
 \label{f:beam-splitter-calibration}%
\end{figure}

Haploscope calibration can be broken down into multiple, discrete steps, listed here.  For a more in-depth examination of each step, please see Massey\cite{Massey:2018:PCOH}.
\begin{enumerate}
	\item{Ensure that the haploscope rests upon a flat, level surface and that each element of the haploscope is also flat and level.  Further, ensure that the two rotating sides of the haploscope are flat, level, and parallel with each other (Figure~\ref{f:haploscope}).} 
	\item{Ensure that the monitors, and all optical components besides the beamsplitters, are mounted such that they are centered along the optical axis of their respective rotating component (Figure~\ref{f:haploscope}).}
	\item{Ensure that the collimation and minimization lenses are positioned on the rails such that they collimate the image from the monitor (Figures~\ref{fig:ray-diagram} and~\ref{f:Diopter}).}
	\item{Ensure that the beam splitters are centered such that they are directly in front of the pivot points of both rotating components.  Further, ensure that the beam splitters are angled, positioned, and tilted so as to ensure that each rotating component's optical axis is reflected through the beamsplitter such that the reflected ray is then parallel to each rail's optical axis and passes through the component pivot point (Figure~\ref{f:beam-splitter-calibration}).} 
    \item{Adjust the haploscope IPD for each new user.  Finally, verify, as much as is possible, that all previously checked conditions remain true (Figure~\ref{f:GazeCond}).}
\end{enumerate} 
Note that introduced errors can be compounded in later steps.

\subsection{Step 1 - Tabletop Setup}
The first steps in building or calibrating an accurate haploscope system are to ensure that the structure the haploscope is built on is rigid and flat, and that it is balanced with respect to gravity.   

Next, a gravity-balanced laser level should be set up and adjusted such that it is parallel to the haploscope table.  With this laser level and its autobalancing feature, we are able to use the ruled holes on the haploscope table to generate a precise coordinate plane to use as a ground truth during haploscope calibration.  This coordinate plane defines an accurate and precise baseline for evaluating haploscope alignment, and so is an important component of calibration.
  
\subsection{Step 2 - Optical Element Mounting}
To mount the optical elements to the sides of the haploscope, it is important that they all be centered on their respective rail, and, thus, the corresponding optical axis, and that they be square with respect to the optical axis.  To do this, the laser level coordinate system and a simple mounting block are used to align the center of each element with the center of the haploscope rail. 

Each rail is rotated outward 90 degrees and securely fastened.  The angle of each rail is verified by ensuring that the laser level is shining straight down the center of the rail, as determined by the mounting block center line.  Once each side is secured, the monitors are placed on the rail and centered by matching the monitor calibration crosshair with the laser level crosshair.

At this point, three known quantities have been established: the laser level defines a coordinate system based on the haploscope table; each optical axis is collinear to this coordinate system; and, finally, the screen crosshair is centered on each rail's optical axis.  With these three known quantities, we can center each optical element individually as we add it to the haploscope.

\subsection{Step 3 - Collimating the Image Source}
Next, it is important that the output from the monitor be collimated; without collimation, it would be untenable to adjust the focal demand of the presented image precisely and accurately, as is required.

To do this, the elements are positioned based on their back focal distance and the well-known thin lens equation, 
\begin{equation}
\frac{1}{f} = \frac{1}{u} + \frac{1}{v} ,
\end{equation}
where $f$ represents the focal demand, $u$ represents the object distance, and $v$ represents the image distance (Figure~\ref{fig:ray-diagram}).  In essence, the minimization lens, a planoconcave lens, takes the output from the monitor and converts it into a smaller, virtual image.  This image is positioned at the planoconvex, or collimating, lens' focal distance and is thus displayed at optical infinity.

To verify that the output of this lens system is correct, we use a dioptometer, a device which allows us to determine if an image is collimated (Figure~\ref{f:Diopter}).  

\subsection{Step 4 - Beamsplitter Calibration}
At this point, all optical components should be positioned appropriately, and the rotating wings returned to infinity vergence.

Next, the beamsplitters must be adjusted until they perfectly match their respective rail's optical axis with a user's modeled optical axis (Figure~\ref{fig:ray-diagram}).  This requires fine-tuned adjustment of the position, rotation, and tilt/yaw/roll of each beamsplitter.  When this calibration step is complete, the laser level defining the coordinate system should be able to connect the modeled eye position and the optical axis for each rail (Figure~\ref{f:beam-splitter-calibration}), aligning a user's theoretical optical axis with the optical axis of the rail and optical elements.  

\subsection{Step 5 - IPD Calibration and Verification}
The final step in calibrating the haploscope for a specific user is adjusting the system for that user's interpupillary distance (IPD) (Figure~\ref{f:GazeCond}).  If the system is not adjusted based on IPD, the user's optical axis will not be aligned with each rail's optical axis, causing noticeable distortion, angular errors, and other problems~\cite{lee:2016:effects}.

One approach to adjusting the system IPD is to set up a calibration target with a distance equal to a potential user's IPD (Figure~\ref{f:IPDCalibrationTarget}).  If two laser levels, separated by the given IPD, can be aligned to either side of the target while also going through the center of each side's optical axis and center of rotation, then the haploscope has been successfully calibrated for that IPD.  If not, the base will have to be adjusted until both lasers bisect the calibration target, the appropriate centerline, and their side's center of rotation (Figure~\ref{f:IPDBigPicture}).  

This methodology also has the notable advantage of verifying the optical element alignment and calibration, which is an important step in guaranteeing an accurate haploscope calibration.  The various other calibration steps, where relevant, should also be re-examined to ensure that they still hold true.

\section{Future Work}

At this point, the haploscope system has been calibrated, as effectively as we are currently able.  However, there are still some important bounds on haploscope accuracy.  Refraction, tracking errors, optical distortions, and measurement errors could all potentially be detrimental to the accuracy of the final calibrated haploscope; minimizing these errors could represent important future work.

\section{Conclusions}

Haploscope-based experiments could range from research into basic depth perception/cue interactions, to designing accommodation-invariant modalities and software, to examining the effects of visual flow on perception, to considering the relationship between cue conflicts and simulation sickness. Research like this could potentially serve to increase our understanding of the field of augmented reality and advance AR applications.  Better understanding of depth perception in AR, for example, might allow the development of previously unimplementable medical applications; accommodation-invariant display technology might improve AR system performance in far-field applications; and a better understanding of simulation sickness could help improve the user experience in AR~\cite{Singh:2013, Cook:2018}.  These sorts of perceptual experiments would certainly make use of the novel affordances offered by haploscopes, and this research, in turn, could help support, develop, and improve the field of augmented reality.

\acknowledgments{%
This material is based upon work supported by the National Science Foundation, under awards IIS-0713609, IIS-1018413, and IIS-1320909, to J. E. Swan II.  This work was conducted at the Institute for Neurocognitive Science and Technology, and the Center for Advanced Vehicular Systems, at Mississippi State University.  We acknowledge Gurjot Singh for the initial design and calibration of the AR haploscope, and Chunya Hua for developing additional calibration methods.} 

\bibliographystyle{abbrv-doi}
\bibliography{references}
\end{document}